\documentclass{article}
\usepackage{pstricks}

\newcommand{\EcoLab}{{\sffamily\slshape
    \mbox{\raisebox{.5ex}{Eco}\hspace{-.4em}{\makebox[.5em]{L}ab}}}}

\title{Going Stupid with \EcoLab}
\author{Russell K. Standish\\
School of Mathematics and Statistics\\
University of New South Wales}

\begin{document}
\maketitle

\begin{abstract}
  In 2005, Railsback et al. proposed a very simple model ({\em Stupid
    Model}) that could be implemented within a couple of hours, and
  later extended to demonstrate the use of common ABM platform
  functionality. They provided implementations of the model in several
  agent based modelling platforms, and compared the platforms for ease
  of implementation of this simple model, and performance.

In this paper, I implement Railsback et al's Stupid Model in the
 \EcoLab{} simulation platform, a C++ based modelling platform,
 demonstrating that it is a feasible platform for these sorts of
 models, and compare the performance of the implementation with
 Repast, Mason and Swarm versions.
\end{abstract}

\section{Introduction}

Newcomers to {\em agent based modelling} (ABM) will be confused by the
variety of different software platforms available to assist in
implementing the models. Very few comparative studies between the
different platforms have been done, as it is a time consuming task
implementing all but the most trivial of models. Furthermore,
familiarity with one platform and programming language will lend an
automatic advantage in any metrics to that platform over other
platforms that the model implementer is less familiar with.

In 2005, Railsback et al.\cite{Railsback-etal06} proposed a very
simple model that could be implemented within a couple of hours, and
later extended to demonstrate the use of common ABM platform
functionality. They gave it the name ``Stupid Model'', partly for fun,
but also to reiterate the recommendation of Grimm and Railsback
\cite{Grimm-Railsback05} that modelling projects should start with a
``ridiculously simplified model''. Railsback et al. implemented their
model across a range of ABM platforms: Objective-C and Java
Swarm\cite{Swarm}, Repast\cite{North-etal06} and
Mason\cite{Luke-etal05} (both pure Java implementations) and Netlogo.
This range of platforms reflects the authors' collective programming
expertise in Objective-C and Java, and with Netlogo having low barrier
of entry (Logo was a popular language for teaching school children in
the 1980s).

\EcoLab{} grew out of a simulation platform supporting a particular
class of model, into a general purpose simulation environment using
C++\cite{Standish-Leow03}. Other C++ agent-based modelling environments
exist, eg SymBioSys\cite{McFadzean94}, but none are as general purpose
as \EcoLab{}. Other general purpose agent based platforms can be used
with C++ models. For instance, with Swarm, C++ code can be linked to
Swarm's objective C library through the shared C language interface,
and C++ code can be linked to Repast's Java library through the Java
Native Interface. However, maintaining the interface code quickly
becomes prohibitive in the face of evolving models, negating much of
the benefits in using a simulation platform in the first place.

With \EcoLab{}, it is possible to have a similar level of
functionality as provided by Swarm or Repast for models implemented in
C++, without the interface maintenance overhead. Additionally,
\EcoLab{} provides features for distributing the computation over
multiple processors in a way that is easier to program than the raw
Message Passing Interface (MPI)\cite{mpiref}. With Railsback et al.'s
Stupid Model specification, the possibility exists for directly
comparing an \EcoLab{} implemented agent based model with other
platforms for both ease of implementation, and execution
performance. Furthermore, the exercise illuminates those parts of
\EcoLab{} requiring improvement.

\subsection{Why C++}

C++\cite{Stroustrup97} is a mature object oriented programming
language of more than 20 years standing. It has been widely adopted in
industry, consequently open source reference compilers, as well as
vendor-tuned optimising compilers exist for most contemporary computer
architectures. Because of this popularity, and the availability of
compilers, C++ has been extensively deployed for scientific computing
since the mid-1990s. In {\em High Performance Computing} (HPC), the extreme
end of scientific computing, the predominant computing language used
for applications is Fortran, with code written in Fortran 77, or
increasingly written using the newer Fortran 90 features. However
C/C++ applications also make up a substantial fraction of the deployed
applications, perhaps as high as 30\%, with C++ standing to C in the
same relationship as Fortran 90 does to Fortran 77, i.e. typically
used as a ``better C''\footnote{These numbers come from a decade of
  personal experience at managing the resource allocation process at a
  High Performance Computing Centre. These general numbers are backed
  up by anecdotal reports from a number of other people I have
  corresponded with}. By contrast, Java\cite{GoslingJava} has made negligible impact in
HPC\footnote{Over the ten years of my personal
  experience, only one project used Java, out of several hundred that
  were mostly C/C++ or Fortran.}. There are several possible reasons
for the lack of Java adoption in high performance computing. Firstly,
most implementations compile to a virtual machine, 
and early Java Virtual Machines (JVMs) had performance
problems. However, more recent JVMs deploy {\em just in time
  compilation}, which closes the performance gap between JVM executed
code and natively compiled code. Secondly, certain language features
missing in Java
(notably operator overloading, and to a lesser extent generic
programming) of C++ (and Fortran 90 for that matter) assist in writing
scientific codes that are closer to the mathematical
specification. However, probably the most significant factor is
time and innate conservatism of scientific programmers. C++ did not
appear significantly in HPC applications until around 15 years after
the language was first developed. With only a decade under its belt,
Java's time as an HPC application language might just be
beginning\cite{JavaGrande}.

However, for agent based simulation, C++ is not a popular choice,
primarily due to its lack of {\em reflection}. Reflection is the
ability to query an object's type information at runtime, and in ABM
systems like Swarm, reflection is used to implement {\em probes}, or
the ability to observe all parts of a running simulation from within a
graphical user interface\cite{Swarm}. However, with Classdesc, an
effective reflection mechanism for C++ is
possible\cite{Madina-Standish01,Standish-Madina06}. \EcoLab{} uses
Classdesc to implement probing, along with automatic checkpointing,
the ability to script the model's initialisation and ongoing
computation, and for distributing agents to exploit any parallel
computing capability. 

\section{Method}\label{method}

In line with Railsback et al.'s\cite{Railsback-etal06} methodology, I
implemented Stupid Model using the current \EcoLab{} release, version
4.D21. This is important to give a sense of the maturity of the
platform. Otherwise, I might have been tempted to fix up any
weaknesses encountered.

I followed the the explicit model specification\cite{StupidSpec} step
by step, referring to the Repast Java implementation on the rare
occasions the specification was ambiguous. Stupid Model consists of
agents called ``Stupid Bugs'' moving around a Cartesian lattice. No
two agents can occupy the same location, so movement involves
selecting a cell within a $9\times9$ Moore neighbourhood, testing whether
the cell is occupied and moving into the cell if empty. The search
procedure is repeated until an empty cell is found. Since different
frameworks potentially use different random number algorithms,
initialised with a different seed, this introduces indeterminism into
model runtimes. In order to reduce the impact of this indeterminism,
the density of agents was chosen to be 0.1 (4000 agents in a
$200\times200$ world) so that the standard deviation of runtimes was
less than 10\% of the mean.

For measuring application performance, I did both GUI runs, and batch
mode runs. In \EcoLab{}, a non-GUI batch run simply involves replacing
the ``GUI'' command from the experiment script, with a call to
``simulate'', and commenting out any graphical calls (plot, histogram and
draw). In Repast, Swarm and Mason, a separate ``BatchSwarm'' needs to
be provided by the programmer, but only the GUI versions of each model
were published by Railsback et al.  For batch measurements, I commented out
the call to addAction that added the display actions. For the Repast
implementation, I changed the batch parameter of
\verb+SimInit::loadModel+ to \verb+true+, and timed the run from the
command line. With the Mason implementation, I again commented out the
display action, and recorded the CPU time so as to discount the delays
introduced by having to click the button. In fact for all platforms,
the reported values are the CPU time. For the Objective C Swarm
version, I modified the code so that the \verb+StupidModelSwarm+ was
directly called from \verb+main()+ rather than indirectly through
\verb+StupidModelObserverSwarm+. 

I chose to measure the versions 10 and 11 of the Stupid
Model. However, the stopping criteria is specified as when the maximum
bug size reaches 100. Since bug growth depends on the availability of
food, which itself is a function of a random number generator call,
and also of the grazing history, this stopping criterion is
indeterministic. For the purposes of inter-framework performance
comparisons, I changed the stopping criterion to be a fixed number of
bug updates (500). 

In version 10 of Stupid Model, bugs will randomly select a cell within
their neighbourhood, and moving to it if the cell is empty, otherwise
repeating the selection process. In version 11, all cells in the
neighbourhood are iterated over, and the bug moves to the empty cell
with the most food.

From version 12, bugs can reproduce and die according to random
dynamics, so the amount of work per update step will depend on the
number of living bugs. Even though these higher version models are
more computationally intensive, run times cannot be compared between
different platforms due to differences in the order that random numbers
are generated. Hence the Stupid 16 measurements reported in table
\ref{execution times} should be taken with a certain amount of
salt. Nevertheless, I verified that all models executed for 1000
steps, and that the number of Stupid Bugs was roughly the same for
each platform (approximately 8-900 after the initial population explosion).

Railsback et al. did not do any performance analysis or tuning. For
C++ code, performance tuning can deliver big performance
improvements. \EcoLab{} can be built with performance counters enabled for the
individual TCL commands, and a single run indicated that the initial
approach used for evaluating the stopping criterion (evaluating the
maximum of the vector of bug sizes in TCL) was very expensive. By
implementing a specialised \verb+max_bugsize()+ (all of 4 lines of C++
code) improved performance by about a factor of four. However, for the
inter-platform performance comparison, the stopping condition was
changed to a fixed number of bug update steps, so this optimisation
makes no difference to the performance benchmarks.

A more detailed performance profile using the standard GNU/Linux
profiling tool {\tt gprof}, indicated that updating the food
availability was a bottleneck, and that cache utilisation could be
improved by laying the data contiguously in memory, which is not the
case when the data is stored as members of a cell object. This
optimisation, which needed some substantial recoding of the model,
improved overall performance by a factor of two for model version 16,
although it only made about a 10\% improvement for version 11. It
should be noted that this optimisation technique should also be
available for the Java and Objective-C platforms, and presumably may
deliver a similar performance boost.

All performance benchmarks were run on a 2GHz Intel Pentium M
processor with 1GB memory running Slackware Linux 10.0. The Java
version used for Repast and Mason was SDK 1.4.2 standard edition. The
compiler used for Swarm and \EcoLab{} was GCC 3.4.3. I also did a
comparison \EcoLab{} run using the Intel C++ compiler 9.0, but this
was more than 50\% slower than the GCC compiled code. This somewhat
surprising result indicates that icc's strength lies in vectorising
loops that access data contiguously to exploit the inbuilt SSE
instructions, but that for more general purpose ABM code, GCC performs
better (at least on Linux!).

The sourcecode for \EcoLab{} Stupid Model is available from the
\EcoLab{} website.\cite{EcoLab}

\section{Results}

Similar to all the platforms reviewed by Railsback et al., \EcoLab{}
proved capable of implementing all functionality for all versions of
Stupid Model. Implementing the first version took longer than any of
the remaining versions, as \EcoLab{} does not provide a ready-to-use
spatial library. Instead it provides a more general library called
{\em Graphcode}\cite{Standish-Madina06}. Graphcode's abstraction is a
network, or graph of objects, with the links between objects
representing data flow. Graphcode can distribute the objects across
multiple processors using the Classdesc serialisation library. A
cellular space such as found in Swarm or Repast will be a set of
objects, each one wired to its neighbours. In such a way, Graphcode
can easily represent Cartesian and hexagonal topologies by the way the
neighbourhoods are wired. However, the only example using Graphcode
provided in the \EcoLab{} was a continuous space example, each cell
holding objects located within a certain region of space. Examples of
models using different sorts of spatial topologies, as well as a few
common cases being supplied as a library would improve the beginner's
experience of \EcoLab{}.

In retrospect, it may have been simpler to implement the spatial class
on top of a standard vector of cells. This would have gotten the
initial model up and running quicker, but limited the model to
sequential usage only. By using Graphcode, we enable parallel
processing capability.

One thing that became clear in this exercise is the need for a smart
reference type. Objects like bugs need a reference to the cell in
which they inhabit, scheduling lists need references to the bugs that
they schedule and so on. Because bugs move from cell to cell, it is
better for the cells to have a reference to the bug it contains (if
any) rather than for the cell to store the bug itself. In C, the only possibility for
references are pointers, which are difficult to serialise properly due
to the fact that C makes no guarantees about whether a pointer is
valid or not. Substantial care is required to ensure that references
remain valid in the event of an object such as a bug being deleted
from the system.  Classdesc accepts a pragma that asserts that a
pointer is either valid or NULL, and whether the pointer chains form
cycles or not to allow serialisation, but it's up to the programmer to
ensure software bugs do not invalidate this assertion. 

C++ also supports static references (eg \verb+int&+), which are
established at the time of the reference's creation, and then
immutable until the reference is destroyed. These references are
always valid, however the lack of dynamic control makes them
unsuitable for agent based simulations where agents may be dropped or
moved, and appropriate references updated. Furthermore static
reference cycles cannot be handled with serialisation at all, since
the serialisation descriptors cannot distinguish an object from its reference.

Whilst it is possible to use \EcoLab{} with a nonserialisable model,
one gives up substantial functionality doing so, including the ability to
checkpoint/restart the model.

What is needed actually is something like Java's reference type, where
objects are created on the heap, and the programmer simply manipulates
references. Once all references to an object have been destroyed,
Java's garbage collector takes care of destroying the object,
reclaiming the memory used.

It is possible to implement something like this in C++, using operator
overloading to give the resulting type the ``look and feel'' of a
pointer. Such types are usually called {\em smart pointers}. The well
known Boost library\cite{Boost} provides a few different versions,
some of which are being considered for inclusion in the C++ standard
library. \EcoLab{} provides the template \verb+ref<T>+, which is
parameterised by the target type of the reference. Unlike the Boost
versions (in which you pass the smart pointer a pointer for it to
control), \verb+ref+ has control over the entire lifecycle of the
object it points to. The first time a \verb+ref+ object is
dereferenced, the target object is created on the heap, and it keeps
track of the number of references to the target object, so that once
all references to are destroyed, so is the target object.

The version of \verb+ref+ supplied in the current \EcoLab{} has a
number of deficiencies, however, most notable of which is that it
doesn't provide any way of testing whether the target object exists or
not. For the purposes of this exercise, I copied the \verb+ref.h+
header file, and added the necessary functionality. This improved
\verb+ref.h+ will be incorporated in future releases of \EcoLab{}.

Agents usually need to refer to the environment, or world in which
they live. In languages like Java or Objective C, this is simply
managed by having the agent store a reference to the world, and/or
cell. However, this will set up a reference cycle which will play
havoc with model serialisation if the serialisation algorithm doesn't
explicitly account for cycles. \EcoLab{} provides a routine that
serialises arbitrary graphs constructed with pointer references.
However, it does not currently support the presence of cycles with the
\verb+ref<>+ data type. With C++, however, there is a simple
workaround. The model is a global variable, and agents can refer to
their cell by holding an index into a container of cells stored within
this global model. This is the approach I have taken with Stupid
Model, and indeed this technique is used in other \EcoLab{} models.
However, if the \verb+ref<>+ data type were extended to support
serialisation of cyclic graphs, the method deployed in Java and
Objective C models can be supported as well.

Line counts are often considered a proxy for the amount of effort a
programmer must expend to implement a problem. Table \ref{line count}
shows the line counts for the 16 different Stupid Model cases for each
of the Railsback implementations, as well as the \EcoLab{}
implementation. The \EcoLab{} implementation also includes two
additional cases, which build upon version 16. The model is
parallelised using \EcoLab{}'s MPI-based parallel processing features,
and finally, the ``field'' optimisation whereby the food data is
stored in contiguous memory. \EcoLab{} and the two Java platforms
seems to need a similar number of lines of code, yet the Swarm
implementation needed up to three times the number. Whilst a factor of two or three in
source line count is not particularly significant, it does indicate
that it takes a bit more effort to implement Swarm models.

\begin{table}
\begin{tabular}{r|rrrr}
Version & Repast & Mason & Obj-C Swarm & \EcoLab{} \\
\hline
1 & 158 & 169 &   578 & 253 \\
2 & 158 & 214 &   622 & 259 \\
3 & 250 & 263 &   865 & 281 \\
4 & 256 &     &   896 & 310 \\
5 & 312 & 296 &   968 & 322 \\
6 & 306 & 362 &  1005 & 338 \\
7 & 359 & 316 &  1070 & 337 \\
8 & 258 & 365 &  1144 & 320 \\
9 & 368 & 369 &  1152 & 336 \\
10 & 381& 383 &  1191 & 352 \\
11 & 391& 409 &  1253 & 358 \\
12 & 497& 494 &  1614 & 416 \\
13 & 484&     &  1636 & 419 \\
14 & 501&     &  1360 & 432 \\
15 & 646& 670 &  1761 & 515 \\
16 & 753& 816 &  2174 & 662 \\
parallel &&   &      & 753  \\       
field &  &   &      & 894  \\
\end{tabular}
\caption{Source code line-counts (as reported by the unix command `wc')
  for the different Stupid Model
  versions. Makefiles are not included (Swarm \& \protect\EcoLab{}), since
  these are fairly boiler plate code, and fairly negligible. \protect\EcoLab{}
  counts include the TCL scripts.}
\label{line count}
\end{table}

\begin{table}
\begin{tabular}{r|rrrr}
Version & Repast & Mason & Obj-C Swarm & \EcoLab{} \\
\hline
10 &     3.5  &  3.4   & 71 & 3.9 \\
11 &     32.7 &  21.3  & 165 & 14.9 \\
16 &     44   &  40.5  & 402 & 1014 \\
field &       &       &     & 67 \\
\end{tabular}
\caption{Execution CPU times (in seconds) for several Stupid Model versions for
  different platforms. Versions 10 and 11 were performed in batch mode
  (no graphical output, no GUI control, Mason excepted), version 16 in
  GUI mode with a 
  plot and histogram. \protect\EcoLab{}'s field version uses raster rather
  than canvas for display, and omits the expensive histogram
  widget. All these figures need considerable qualification (see text).}
\label{execution times}
\end{table}

In table \ref{execution times}, execution times for various stupid
model versions is reported. As described in \S\ref{method}, versions
10 \& 11 were run in batch mode with as much graphical output turned
off as possible. The Java versions performed slightly better for
version 10, and the C++ version did better on version 11. However,
given the possible range of implementation strategies, one should not read
too much into this, except that the myth of Java being slow relative
to C++ should be now be firmly laid to rest. The result is broadly in
line with other observations that Java implementations tend to be
within a factor of 2 of natively compiled
applications\cite{Boisvert-etal01,Lewis-Neumann03}. The results for
Swarm though confirm Railsback et al's the observation that Objective
C performance lags that of the Java (and also now C++)
versions. Unfortunately, my knowledge of Objective-C and Swarm
internals is not up to the task of explaining this result.
 
In version 16, the full graphical version of the model was run. This
included a display of the space, a plot of the number of bugs and a
histogram of bug sizes. It should be noted that the Mason
implementation lacked the plot and histogram, apparently because this
functionality is absent within the Mason
toolkit\cite{Railsback-etal06} itself, but provided by 3rd party
add-ons. One thing that stands out is the slowness of \EcoLab{}. The
TCL-based plotting widgets used in \EcoLab{} (also used in Swarm) are
slow relative to the equivalent Java offerings. Furthermore, this
benchmark displays the space environment using a canvas, which is a
high level drawing tool with roughly the same sort of functionality as
a standard drawing application (eg. the drawing application in
OpenOffice or Xfig). The bugs, predators and empty cells are rendered
as coloured squares. The other platforms provide dedicated raster
objects for rendering spatial displays. In the ``field'' version of
Stupid Model, instead of representing the model's objects as squares, a
single pixmap object is created on the canvas and manipulated through
low level Tk library calls. This amounts to about 40 lines of code,
and improves the display performance dramatically. The result listed
under the row ``field'' also omits the expensive histogram
functionality (but still displayed the plot of bug numbers).

\section{Parallel implementation}

Having put the extra work into building the space class on top of
Graphcode rather than using a simple vector, it raises the question of
whether Stupid Model can be effectively parallelised.

The first thing that becomes apparent is that Stupid Model as
specified is inherently sequential. Two bugs are not allowed to occupy
the same spatial location, and movement into a location is performed
on a first come first served basis. Since the order in which bugs
perform their update move is randomised, the obvious parallel
generalisation in a shared memory context is to use locks to prevent
two bugs on different processors simultaneously moving to the same
location. However, \EcoLab{} is designed for use with distributed
parallel systems, and obtaining the state of a cell located on a remote
processor is expensive. In fact, in the MPI transport layer used by
\EcoLab{}, such functionality is only supported by ``one-sided''
communications of MPI 2, a relatively new feature that is not well
supported and typically poorly implemented. Instead, the recommended
approach in \EcoLab{} is to have separate communication and
computation phases, with a snapshot of neighbouring data at the
previous timestep supplied to each processor during the communication
phase.

As Stupid Model is a pedagogical model, there is no one right answer
as to respecifying the model for parallelisation.  Perhaps the most
obvious approach would be to allow multiple bugs to share a single
location within the space. This would certainly simplify the code, as
additional logic was required to enforce the one-bug-per-location
requirement. However, in the spirit of adventure, I propose the
following protocol for allowing bugs to migrate from one processor to
the next, whilst maintaining the one-bug-per-location property. As in
the sequential algorithm, bugs examine their neighbourhood, and choose
the cell with the highest food resource as a destination. If the
destination lies on the current processor, and the cell is empty, the
bug is free to move. If the destination is remote, however, the bug's
desire to move to a remote cell is lodged with an emigration
register. Then after all bugs have performed their move, the
emigration register is passed to the remote processor, which approves
or denies the request depending on whether the destination is already
occupied, or an immigration request has already been allowed. The
immigration approval list is passed back to the requesting processor,
and approved bugs are migrated between processors. The remaining
bugs do not move.

I coded this solution into the \verb+stupid-parallel+ version, and
also the field optimised version \verb+stupid-field+. None of the
other versions are parallel aware code --- building them and running
them in parallel will only result in the model running on processor 0,
with the remaining processors idle.

With the \verb+stupid-parallel+ version, it became immediately clear
that the \verb+Prepare_Neighbours()+ step dominated the
calculation. This highlighted a hitherto unsuspected source of
inefficiency in Graphcode's \verb+Prepare_Neighbours()+ method. To
build the list of neighbours to transmit, Graphcode loops over the
neighbours of local cells, adding to the list any remote neighbour
found. However, this leads to many duplicates, as one cell may be the
neighbour of many other cells --- for the Stupid Model case, each cell
in the transfer list will be duplicated 36 times. In a more common von
Neumann neighbourhood of radius 1 there is no duplication, and in the Moore
neighbourhood of radius 1 the duplication is only 3 times. In choosing
a Moore neighbourhood of radius 4 for their Stupid Model, Railsback et
al. unwittingly made this inefficiency blatant.

However, even with this inefficiency corrected,
\verb+Prepare_Neighbours()+ is still an expensive overhead. The
example problem I tested was the same $200\times200$ spatial grid,
and so $2\times200\times4\times N_p$ cells need to be transferred each
time step ($N_p>1$ being the number of processors). This overhead can
be amortised by increasing the problem size.

In the \verb+stupid-field+ case, the \verb+food_available+ data is not
stored in the cell, but in the additional field data structure, so is
not transferred with the cell data during the
\verb+Prepare_Neighbours()+ step. In fact, only the food data has any
affect on bug movement, so \verb+Prepare_Neighbours()+ is eliminated
altogether. In the \verb+stupid-field+ version of the model, we do not
transfer the food data, but duplicate the update calculation on the
overlap area between two processors. A single
\verb+Prepare_Neighbours()+ step is done at the beginning of the model
run to ensure access to the food data.

\begin{figure}
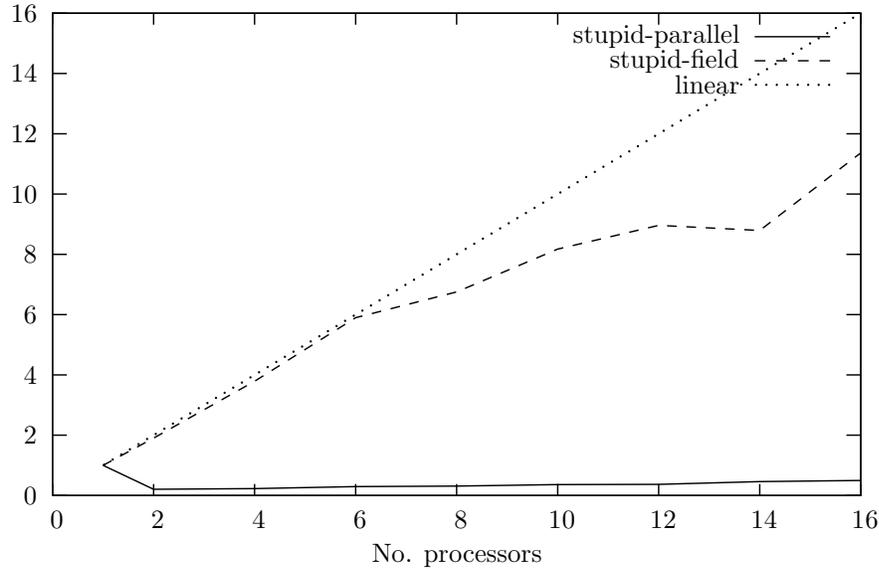


% Define new PST objects, if not already defined
\ifx\PSTloaded\undefined
\def\PSTloaded{t}
\psset{arrowsize=.01 3.2 1.4 .3}
\psset{dotsize=.01}
\catcode`@=11

\newpsobject{PST@Border}{psline}{linewidth=.0015,linestyle=solid}
\newpsobject{PST@Axes}{psline}{linewidth=.0015,linestyle=dotted,dotsep=.004}
\newpsobject{PST@Solid}{psline}{linewidth=.0015,linestyle=solid}
\newpsobject{PST@Dashed}{psline}{linewidth=.0015,linestyle=dashed,dash=.01 .01}
\newpsobject{PST@Dotted}{psline}{linewidth=.0025,linestyle=dotted,dotsep=.008}
\newpsobject{PST@LongDash}{psline}{linewidth=.0015,linestyle=dashed,dash=.02 .01}
\newpsobject{PST@Diamond}{psdots}{linewidth=.001,linestyle=solid,dotstyle=square,dotangle=45}
\newpsobject{PST@Filldiamond}{psdots}{linewidth=.001,linestyle=solid,dotstyle=square*,dotangle=45}
\newpsobject{PST@Cross}{psdots}{linewidth=.001,linestyle=solid,dotstyle=+,dotangle=45}
\newpsobject{PST@Plus}{psdots}{linewidth=.001,linestyle=solid,dotstyle=+}
\newpsobject{PST@Square}{psdots}{linewidth=.001,linestyle=solid,dotstyle=square}
\newpsobject{PST@Circle}{psdots}{linewidth=.001,linestyle=solid,dotstyle=o}
\newpsobject{PST@Triangle}{psdots}{linewidth=.001,linestyle=solid,dotstyle=triangle}
\newpsobject{PST@Pentagon}{psdots}{linewidth=.001,linestyle=solid,dotstyle=pentagon}
\newpsobject{PST@Fillsquare}{psdots}{linewidth=.001,linestyle=solid,dotstyle=square*}
\newpsobject{PST@Fillcircle}{psdots}{linewidth=.001,linestyle=solid,dotstyle=*}
\newpsobject{PST@Filltriangle}{psdots}{linewidth=.001,linestyle=solid,dotstyle=triangle*}
\newpsobject{PST@Fillpentagon}{psdots}{linewidth=.001,linestyle=solid,dotstyle=pentagon*}
\newpsobject{PST@Arrow}{psline}{linewidth=.001,linestyle=solid}
\catcode`@=12

\fi
\psset{unit=5.0in,xunit=5.0in,yunit=3.0in}
\pspicture(0.000000,0.000000)(1.000000,1.000000)
\ifx\nofigs\undefined
\catcode`@=11

\PST@Border(0.1010,0.1260)
(0.1160,0.1260)

\PST@Border(0.9470,0.1260)
(0.9320,0.1260)

\rput[r](0.0850,0.1260){ 0}
\PST@Border(0.1010,0.2313)
(0.1160,0.2313)

\PST@Border(0.9470,0.2313)
(0.9320,0.2313)

\rput[r](0.0850,0.2313){ 2}
\PST@Border(0.1010,0.3365)
(0.1160,0.3365)

\PST@Border(0.9470,0.3365)
(0.9320,0.3365)

\rput[r](0.0850,0.3365){ 4}
\PST@Border(0.1010,0.4418)
(0.1160,0.4418)

\PST@Border(0.9470,0.4418)
(0.9320,0.4418)

\rput[r](0.0850,0.4418){ 6}
\PST@Border(0.1010,0.5470)
(0.1160,0.5470)

\PST@Border(0.9470,0.5470)
(0.9320,0.5470)

\rput[r](0.0850,0.5470){ 8}
\PST@Border(0.1010,0.6523)
(0.1160,0.6523)

\PST@Border(0.9470,0.6523)
(0.9320,0.6523)

\rput[r](0.0850,0.6523){ 10}
\PST@Border(0.1010,0.7575)
(0.1160,0.7575)

\PST@Border(0.9470,0.7575)
(0.9320,0.7575)

\rput[r](0.0850,0.7575){ 12}
\PST@Border(0.1010,0.8628)
(0.1160,0.8628)

\PST@Border(0.9470,0.8628)
(0.9320,0.8628)

\rput[r](0.0850,0.8628){ 14}
\PST@Border(0.1010,0.9680)
(0.1160,0.9680)

\PST@Border(0.9470,0.9680)
(0.9320,0.9680)

\rput[r](0.0850,0.9680){ 16}
\PST@Border(0.1010,0.1260)
(0.1010,0.1460)

\PST@Border(0.1010,0.9680)
(0.1010,0.9480)

\rput(0.1010,0.0840){ 0}
\PST@Border(0.2068,0.1260)
(0.2068,0.1460)

\PST@Border(0.2068,0.9680)
(0.2068,0.9480)

\rput(0.2068,0.0840){ 2}
\PST@Border(0.3125,0.1260)
(0.3125,0.1460)

\PST@Border(0.3125,0.9680)
(0.3125,0.9480)

\rput(0.3125,0.0840){ 4}
\PST@Border(0.4183,0.1260)
(0.4183,0.1460)

\PST@Border(0.4183,0.9680)
(0.4183,0.9480)

\rput(0.4183,0.0840){ 6}
\PST@Border(0.5240,0.1260)
(0.5240,0.1460)

\PST@Border(0.5240,0.9680)
(0.5240,0.9480)

\rput(0.5240,0.0840){ 8}
\PST@Border(0.6298,0.1260)
(0.6298,0.1460)

\PST@Border(0.6298,0.9680)
(0.6298,0.9480)

\rput(0.6298,0.0840){ 10}
\PST@Border(0.7355,0.1260)
(0.7355,0.1460)

\PST@Border(0.7355,0.9680)
(0.7355,0.9480)

\rput(0.7355,0.0840){ 12}
\PST@Border(0.8413,0.1260)
(0.8413,0.1460)

\PST@Border(0.8413,0.9680)
(0.8413,0.9480)

\rput(0.8413,0.0840){ 14}
\PST@Border(0.9470,0.1260)
(0.9470,0.1460)

\PST@Border(0.9470,0.9680)
(0.9470,0.9480)

\rput(0.9470,0.0840){ 16}
\PST@Border(0.1010,0.1260)
(0.9470,0.1260)
(0.9470,0.9680)
(0.1010,0.9680)
(0.1010,0.1260)

\rput(0.5240,0.0210){No. processors}
\rput[r](0.8200,0.9270){stupid-parallel}
\PST@Solid(0.8360,0.9270)
(0.9150,0.9270)

\PST@Solid(0.1539,0.1786)
(0.1539,0.1786)
(0.2068,0.1368)
(0.3125,0.1380)
(0.4183,0.1415)
(0.5240,0.1424)
(0.6298,0.1450)
(0.7355,0.1454)
(0.8413,0.1502)
(0.9470,0.1522)

\rput[r](0.8200,0.8850){stupid-field}
\PST@Dashed(0.8360,0.8850)
(0.9150,0.8850)

\PST@Dashed(0.1539,0.1786)
(0.1539,0.1786)
(0.2068,0.2261)
(0.3125,0.3254)
(0.4183,0.4364)
(0.5240,0.4814)
(0.6298,0.5562)
(0.7355,0.5976)
(0.8413,0.5887)
(0.9470,0.7241)

\rput[r](0.8200,0.8430){linear}
\PST@Dotted(0.8360,0.8430)
(0.9150,0.8430)

\PST@Dotted(0.1539,0.1786)
(0.1539,0.1786)
(0.1619,0.1866)
(0.1699,0.1946)
(0.1779,0.2025)
(0.1859,0.2105)
(0.1939,0.2185)
(0.2019,0.2265)
(0.2100,0.2344)
(0.2180,0.2424)
(0.2260,0.2504)
(0.2340,0.2584)
(0.2420,0.2663)
(0.2500,0.2743)
(0.2580,0.2823)
(0.2660,0.2903)
(0.2740,0.2982)
(0.2821,0.3062)
(0.2901,0.3142)
(0.2981,0.3221)
(0.3061,0.3301)
(0.3141,0.3381)
(0.3221,0.3461)
(0.3301,0.3540)
(0.3381,0.3620)
(0.3461,0.3700)
(0.3542,0.3780)
(0.3622,0.3859)
(0.3702,0.3939)
(0.3782,0.4019)
(0.3862,0.4099)
(0.3942,0.4178)
(0.4022,0.4258)
(0.4102,0.4338)
(0.4183,0.4418)
(0.4263,0.4497)
(0.4343,0.4577)
(0.4423,0.4657)
(0.4503,0.4736)
(0.4583,0.4816)
(0.4663,0.4896)
(0.4743,0.4976)
(0.4823,0.5055)
(0.4904,0.5135)
(0.4984,0.5215)
(0.5064,0.5295)
(0.5144,0.5374)
(0.5224,0.5454)
(0.5304,0.5534)
(0.5384,0.5614)
(0.5464,0.5693)
(0.5544,0.5773)
(0.5625,0.5853)
(0.5705,0.5932)
(0.5785,0.6012)
(0.5865,0.6092)
(0.5945,0.6172)
(0.6025,0.6251)
(0.6105,0.6331)
(0.6185,0.6411)
(0.6265,0.6491)
(0.6346,0.6570)
(0.6426,0.6650)
(0.6506,0.6730)
(0.6586,0.6810)
(0.6666,0.6889)
(0.6746,0.6969)
(0.6826,0.7049)
(0.6906,0.7128)
(0.6986,0.7208)
(0.7067,0.7288)
(0.7147,0.7368)
(0.7227,0.7447)
(0.7307,0.7527)
(0.7387,0.7607)
(0.7467,0.7687)
(0.7547,0.7766)
(0.7627,0.7846)
(0.7708,0.7926)
(0.7788,0.8006)
(0.7868,0.8085)
(0.7948,0.8165)
(0.8028,0.8245)
(0.8108,0.8325)
(0.8188,0.8404)
(0.8268,0.8484)
(0.8348,0.8564)
(0.8429,0.8643)
(0.8509,0.8723)
(0.8589,0.8803)
(0.8669,0.8883)
(0.8749,0.8962)
(0.8829,0.9042)
(0.8909,0.9122)
(0.8989,0.9202)
(0.9069,0.9281)
(0.9150,0.9361)
(0.9230,0.9441)
(0.9310,0.9521)
(0.9390,0.9600)
(0.9470,0.9680)

\PST@Border(0.1010,0.1260)
(0.9470,0.1260)
(0.9470,0.9680)
(0.1010,0.9680)
(0.1010,0.1260)

\catcode`@=12
\fi
\endpspicture
\caption{Speedup curves for {\tt stupid-parallel} and
{\tt stupid-field} for a $200\times200$ grid with 4000 stupid bugs
moving and growing. Bug reproduction and mortality as well as
predation have been turned off. At no stage does {\tt stupid-parallel}
run as fast in parallel as it does sequentially, due to the overheads
of the {\tt Prepare\_Neighbours()} step.}
\label{speedup}
\end{figure}

Figure \ref{speedup} shows the speedup curve for both the
\verb+stupid-parallel+ and \verb+stupid-field+ model, for the same
input script used for the \verb+stupid10+ and \verb+stupid11+
benchmarks reported in table \ref{execution times}.

The parallel computing experiements were performed on Linux cluster
(Beowulf style) with dual 3GHz Pentium 4 Xeon nodes connected via
Gigabit Ethernet. Each node has 2GB of memory.

\section{Conclusion}

The aim of this study was to answer the following questions:
\begin{itemize}
\item is \EcoLab{} suitable for the sorts of agent based models that
  other more well known platforms are used for
\item what performance advantages, if any, does the use of C++
  provide
\item what deficiencies are present in \EcoLab{}
\end{itemize}

Stupid Model is a nontrivial, yet fairly simple agent based model that
could be implemented without an excessive amount of programming. 
\EcoLab{} has shown itself to be capable of implementing Stupid Model
with about the same sort of effort reported by developers of Repast
and Mason versions of the model, and was implemented in around the
same number of lines of code. Furthermore, performance was on a par
with these Java-based platforms.

The main deficiencies encountered were:
\begin{itemize}
\item A lack of specialised space library, or library of examples in
  the use of Graphcode for implementing spaces.
\item A lack of a simple raster object for displaying spaces. The provided canvas
  functionality is very slow
\item GUI functionality is slow compared with the Java-based
  functionality
\item the smart pointer template \verb+ref+ needs to be improved
\end{itemize}

For addressing the space library issue, I will start with implementing
a few well known ABM models to build up a library of practice. Where
code appears in common, this can be refactored into a library.

To address the GUI performance, a possible future strategy is to develop a
Classdesc C++/Java interface to enable C++ coded \EcoLab{} models to
run under a Java framework such as Repast. A similar strategy was
investigated integrating C++ and Objective C using Classdesc to look
at Swarm integration, however it never found practical use and is no
longer being maintained\cite{Leow-Standish03}.  The feasibility of doing 
this with a Java platform will be the subject of future work.

%\bibliographystyle{plain}
%\bibliography{rus}

\end{document}